\documentclass[letterpaper]{article} 

\usepackage{amssymb,amsmath}
\usepackage{graphicx}
\usepackage{authblk}
 \pdfoutput=1

\title{Equations for Solar Tracking}

\author{Alexis Merlaud}
\author{Martine De Mazi\`ere}
\author{Christian Hermans}
\affil{Belgian Institute for Space Aeronomy, Avenue
Circulaire 3, 1180 Brussels, Belgium}
\author{Alain Cornet}
\affil{Institute of Condensed Matter and Nanosciences, UCL, Chemin du Cyclotron 2, 1348, Louvain-La-Neuve, Belgium}

\begin{document}

\maketitle

\begin{abstract}
Direct sunlight absorption by trace gases can be used to
quantify them and investigate atmospheric chemistry. In such
experiments, the main optical apparatus is often a grating or a
Fourier transform spectrometer. A solar tracker based on motorized
rotating mirrors is commonly used to direct the light along the
spectrometer axis, correcting for the apparent rotation of the Sun.
Calculating the Sun azimuth and altitude for a given time and
location can be achieved with high accuracy but different sources of
angular offsets appear in practice when positioning the mirrors. A
feedback on the motors, using a light position sensor close to the
spectrometer, is almost always needed. This paper aims to gather the
main geometrical formulas necessary for the use of a widely used
kind of solar tracker, based on two 45$^{\circ}$ mirrors in
altazimuthal set-up with a light sensor on the spectrometer, and to
illustrate them with a tracker developed by our group for
atmospheric research.   
\end{abstract}

\section{Introduction}

Spectroscopic analyses of direct incident sunlight are commonly used
in atmospheric research. Such experiments make use of the Sun as a
light source to quantify molecular absorptions in the atmosphere and
then retrieve trace gas abundances. Stratospheric
ozone~\cite{barret2002} and greenhouse gases~\cite{demaziere_2005}
are routinely measured with this technique from ground-based Fourier
transform infrared (FTIR) spectrometers, e.g.,~within the Network
for the Detection of Atmospheric Composition Change (NDACC,
http://www.ndacc.org/). In the UV-visible range, light
scattering is more important and enables spectroscopic studies of
the atmosphere in other geometries such as zenith
measurements~\cite{mvr_1998}. However, direct
sunlight is also used~\cite{cageao_2001,wang-2010}, its unique and
unambiguous light path making it advantageous for some
applications~\cite{Spinei2011}. Beside the spectrometer, the main
part of the involved apparatus in direct sunlight spectrometry is
the solar tracker, required to compensate for the Sun's diurnal
motion.

\begin{figure}[ht]
\centering
\includegraphics[width=80mm]{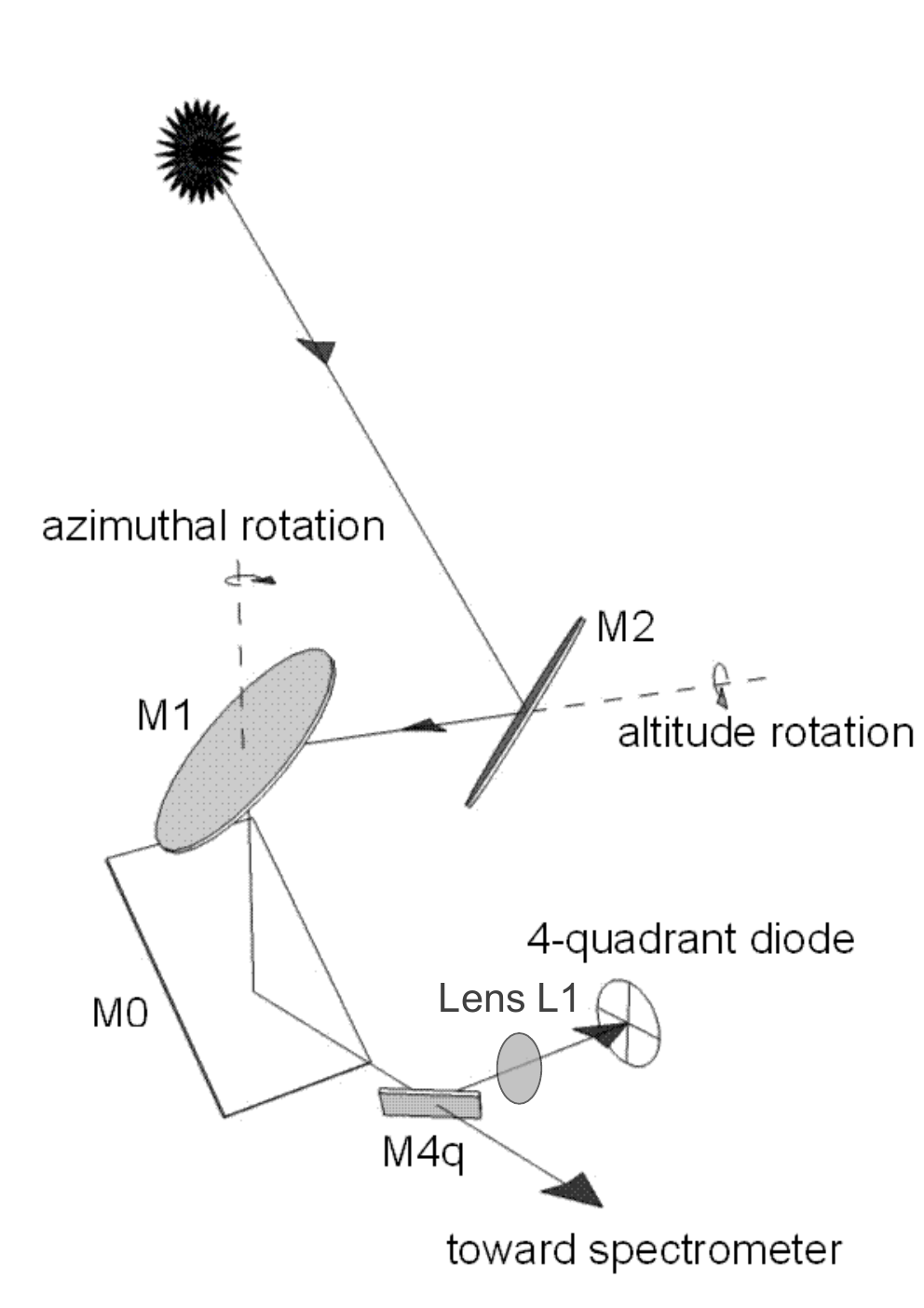}
\caption{Geometrical setup of the considered solar tracker, using
two 45-degree mirrors, M1 and M2, rotating along orthogonal axes.
Mirror M0 directs the Sun light into a spectrometer. A fraction of
the light beam is deflected toward a 4-quadrant photodiode enabling
a \mbox {closed-loop} control of the mirrors position.}
\label{fig:principle}
\end{figure}

Several kinds of trackers, sometimes referred to as
heliostats, are used for atmospheric spectrometry, based on setups
of one or several rotating mirrors. Some of them are equatorially
mounted, like in Table Mountain Facility~\cite{cageao_2001} or
Harestua~\cite{galle_1999}. In this case, one rotational axis is
parallel to the Earth's axis. It enables a high tracking accuracy
without a computer, since only one axis has to be driven at the
Earth's rotation speed. To our knowledge, it is the only setup
working without feedback on the Sun's position. On the other hand,
equatorial mounts are large, need to be aligned accurately and their
mechanical design is difficult. Most of the trackers used today are
controlled by a computer enabling remote operation
and automation. The computer first
calculates the Sun position, moves the mirrors to
point to the Sun and then controls these mirrors to
optimize the signal on some kind of light sensor.
For some trackers, the light sensor is attached to
the moving part, whether it is a single mirror~\cite{wiacek_2007} or
a mount of two mirrors~\cite{neefs_2007}. Compared to the solution
presented below, the retroaction is simplified. The drawback is that
the tracking is done some meters away from the spectrometer and is
thus less accurate and stable.

Figure~\ref{fig:principle} shows a popular altazimuthal tracker
design. It consists of two elliptical mirrors held in 45 degrees
relative to the vertical, facing each other (M1 and M2). Both M1 and
M2 rotate along the azimuthal axis and M2 rotates as well around a
horizontal axis (altitude direction). M0, and possibly other fixed
mirrors, direct the light beam into the spectrometer optical axis. A
4-quadrant photodiode is used as a position sensor for a closed-loop
control of the mirrors position once their positioning towards the
Sun has been set with enough accuracy, \textit{i.e.},~once the Sun's
image is visible by the photodiode. This altazimuthal setup is used
with FTIR systems, e.g.,~in Kiruna~\cite{Huster} and Park Falls,
Wisconsin~\cite{washenfelder_2006}; it has been installed in
Harestua to replace the equatorially mounted
system~\cite{merlaud2006}. Compact versions have also been developed
for field campaigns~\cite{merlaud2004,Cordenier}. A commercial
version is sold by Bruker to be installed on their FTIR
spectrometers~\cite{geibel-2010}. A recent progress in the pointing
accuracy has been reported~\cite{Gisi_2011}, replacing the
traditional quadrant diode with a CCD camera, but the problems
discussed hereafter remain the same.

Because developing a solar tracker is typically a master's thesis
work~\cite{Huster,Cordenier,merlaud2004}, technical implementations
are difficult to access in the literature. Some more information is
available about the systems used in solar energy applications but
their geometries
differ~\cite{Chong_2009,Chong_2009b,guo_2011,wei_2011}. Someone
building a Sun tracker can quickly find ephemeris calculations in
many programming languages, but other issues arise quickly. It is
first necessary to characterize the field-of-view (FOV) of the
4-quadrant diode in the considered optical design. This serves two
purposes: determining the accuracy needed for the ephemeris's
algorithm and making sure this FOV is larger than the Sun's apparent
diameter (9 mrad). This last point is important to track constantly
the center of the Sun, which reduces the uncertainties in the air
mass factor and avoids Doppler shifts on the edges
of the Sun (\cite{Gisi_2011}). A second problem lies in the
correction of the tracker orientation compared to the
altazimuthal system in which the ephemeris is
given, necessary for the calculated mode if the base of the solar
tracker is not leveled. Thirdly, the relationship between the
quadrant signal and the correction to apply on the mirrors positions
depends on the tracker position itself~\cite{Gisi_2011}.
Understanding this relationship is compulsory to achieve a smooth
tracking. This article deals with these three problems successively.

\section{Theoretical Basis}
\subsection{Ephemeris Accuracy and Field of View of the 4-Qua\-drant Diode}
\label{sec:fov}

Calculating the Sun position in the sky given the time of
observation and the geographical coordinates is well documented. A
reference algorithm is given by Jean Meeus
in~\cite{meeus_astronomical_1998}, for which
C(\cite{reda_solar_2008}) or Matlab \footnote{http://www.mathworks.com/matlabcentral/fileexchange/4605-sunposition-m} versions are available. This accuracy degrades with larger zenithal angle due to atmospheric
refraction, which depends on local meteorological conditions
(for an accurate and wavelength dependent refraction
formula, see~\cite{Ciddor96}). Irregular
variations in Earth rotation also limit the accuracy of ephemerids
independently of refraction. On the other hand, the absolute
accuracy of commercial rotation stages used in Sun trackers,
e.g.,~Newport RV-160, is only 0.01$^{\circ}$. This reduces the
interest of using an accurate but complex algorithm for ephemeris's
calculation, which is anyway not necessary providing a closed-loop
control is performed on the mirrors' position. In this case, the
lowest acceptable accuracy is thus determined by the field of view
of the 4-quadrant diode: once the Sun image hits the quadrant, the
tracking can be performed in closed-loop.

\begin{figure}[ht]
\centering
\includegraphics[width=.95\textwidth]{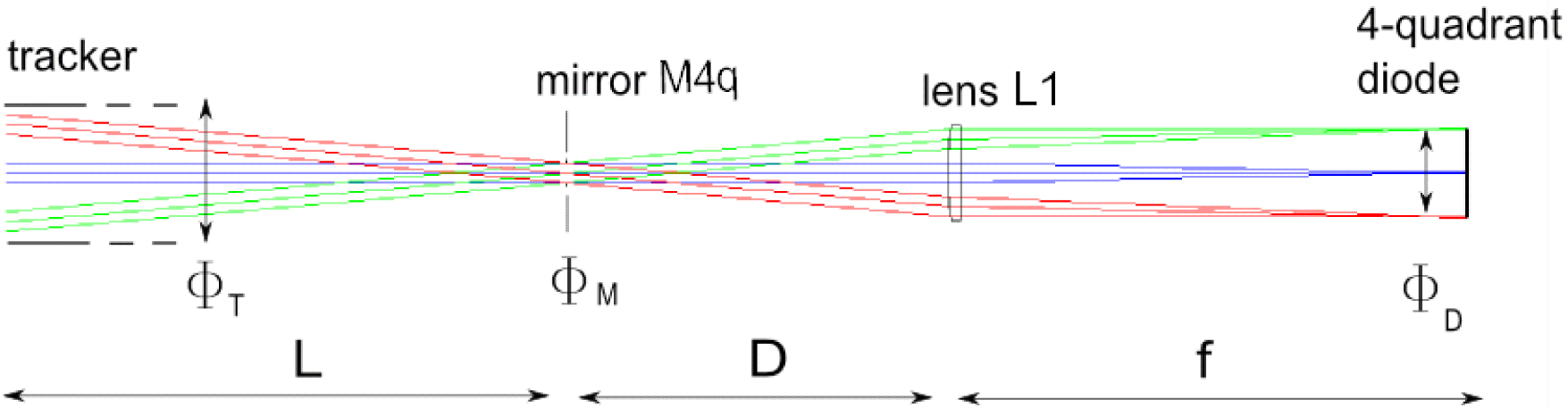}
\caption{Optical scheme from the tracker mirrors to the 4-quadrant
diode. The field of view seen by the diode depends on the different
aperture sizes and path lengths. The green and red
beams represent the Sun light path when the tracker is not perfectly
aligned on the Sun.} \label{fig:fov4q}
\end{figure}

Figure~\ref{fig:fov4q} shows the typical optical scheme between the
tracker and the quadrant diode, the optical axis has been aligned
for the sake of clarity. The distance L is measured from the first
mirror which reflects the sunlight, \textit{i.e.},~M2 in
Figure~\ref{fig:fov4q}. Close to the spectrometer, a part of the
beam from the tracker, with diameter $\Phi_T$ corresponding to the
small axis of M1 and M2 on Figure~\ref{fig:fov4q}, is deflected by a
mirror of diameter $\Phi_M$ to a lens which focuses the beam onto
the quadrant. The maximum field-of-view seen by the diode
($FOV_{1}$) depends on the focal length of the lens ($f$) and on the
diameter of the quadrant ($\Phi_D$), according to $FOV_{1} =
\arctan{\frac{\Phi_D}{f}}$. The lens L1 is unlikely to reduce the
FOV assuming its size superior to the mirror's one. Indeed, the beam
is parallel before the lens which implies that distance D can be
reduced if necessary. The tracker aperture, in our
case defined by the azimuthal stage free aperture, is more
important, especially since the diameter of a rotation stage is
limited and the distance between the tracker and the deflecting
mirror (M4) depends on the observatory configuration. The mirror's
FOV due to the tracker is $FOV_{2} = \arctan{\frac{\Phi_T}{L}}$.

\begin{figure}[ht]
\centering
\includegraphics[width=.8\textwidth]{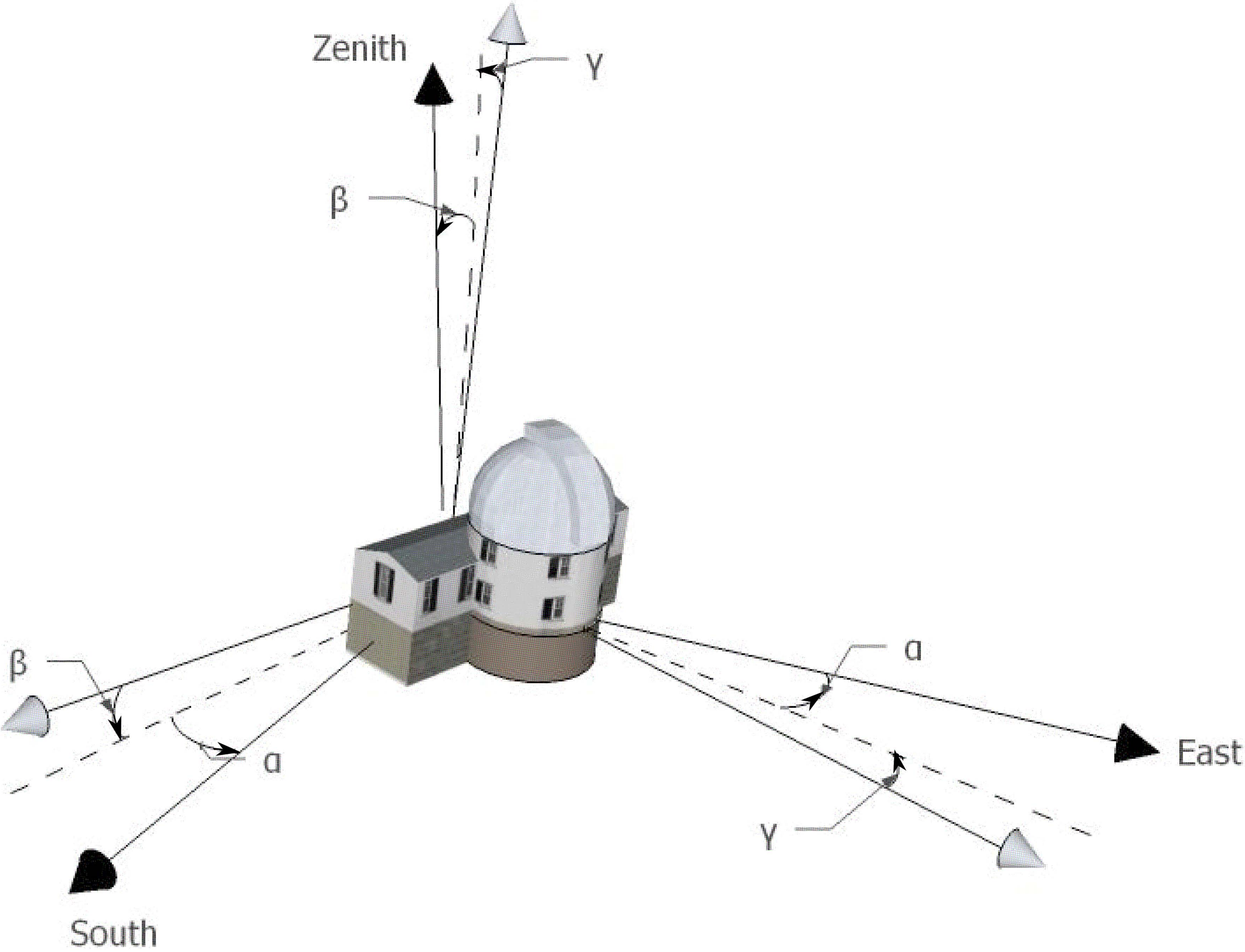}
\caption{Illustration of the Euler Angles of an observatory compared
to the altazimuthal coordinate system.}
\label{fig:euler}
\end{figure}

Considering our set-up in Brussels, which is a typical FTIR station,
the tracker's mirrors are 10 cm ($\Phi_T$) wide and the optical path
to the spectrometer ($L$) 5 m long, which leads to a $FOV_{2}$ of 20
mrad. On the other hand, the quadrant diameter ($\Phi_D$) is 6 mm
while the focal length of the lens (($f$)) is 200 mm, which gives 30
mrad for $FOV_{1}$. The actual FOV is the minimum, 20 mrad, limited
by the tracker size. This value is superior to the apparent diameter
of the Solar disk which is important to track the center of the Sun.
A simple algorithm can achieve such an accuracy for the ephemeris
calculation, like the one given in the appendix. It
is however necessary to take into account the orientation of the
tracker, which can lead to pointing errors superior to the FOV.

\subsection{Correcting the Tracker Orientation}
\label{sec:euler}

One source of error, when pointing to the calculated Sun position,
is the orientation of the baseplate of the tracker compared to the
altazimuthal system. Depending on the observatory configuration, it
may be difficult or impossible to align accurately the tracker along
the North-South direction. If it was the only problem, the remaining
constant offset could be simply fitted and added in the calculated
azimuth. However, because the baseplate is never completely leveled
either, other offsets are added to the calculated positions,
affecting both azimuth and elevation in a way that depends on the
pointing direction of the tracker. Some tracker uses
an active search method to solve this problem. In practice they
reach the calculated position and achieve spiral motion around this
point to set the sun spot in the field-of-view of the sensor.
Misalignment effect can on the other hand be taken into account in
the calculation requires determining the Euler angles of the
observatory and the tracker baseplate, respectively, compared to the
azimuthal system. We discuss the Euler angles before describing our
way to determine them. For the sake of simplicity, we only mention
the observatory in the following, considering that the baseplate
to be part of it.

Euler angles of an observatory may be seen, as in
Figure~\ref{fig:euler}, as consecutive rotations around three
orthogonal axes needed to account for the pitch, roll and yaw of
this observatory. Converting the solar altitude and azimuth to the
observatory frame requires thus to compute the multiplication of
three rotation matrices along the different axes,
\textit{i.e.},~$R_x(\gamma)$, $R_y(\beta)$ and $R_z(\alpha)$.
The resulting matrix $M_{offset}$ expresses the
transformation of coordinates due to the Euler angles.
\begin{align}
\lefteqn{M_{offsets}  = } \\ &
\begin{bmatrix}
1 & 0 & 0   \\
0 & \cos{\gamma} & -\sin{\gamma} \\
0 & \sin{\gamma} & \cos{\gamma}
\end{bmatrix}
\times
 \begin{bmatrix}
\cos{\beta} & 0 & \sin{\beta} \\
0 & 1 & 0 \\
-\sin{\beta} & 0 & \cos{\beta} \end{bmatrix} \times
 \begin{bmatrix}
\cos{\alpha} & -\sin{\alpha} & 0 \\
\sin{\alpha} & \cos{\alpha} & 0 \\
0 & 0 & 1 \end{bmatrix}  \nonumber
\end{align}
The calculations leads to:
\begin{align}
\lefteqn{M_{offsets} =  }\\ &
\begin{bmatrix}
\cos{\alpha}\cos{\beta} & -\sin{\alpha}\cos{\beta} & \sin{\beta} \\
\cos{\alpha}\sin{\beta}\sin{\gamma}+\sin{\alpha}\cos(\gamma) & \cos{\alpha}\cos{\gamma}-\sin{\alpha}\sin{\beta}\sin{\gamma} & -\cos{\beta}\sin{\gamma} \\
\sin{\alpha}\sin{\gamma} -\cos{\alpha}\sin{\beta}\cos{\gamma} &
\cos{\alpha}\sin{\gamma}+\sin{\alpha}\sin{\beta}\cos{\gamma} &
\cos{\beta}\cos{\gamma} \end{bmatrix} \nonumber
\end{align}

In Cartesian coordinates, the unit vector ($x_t$,$y_t$,$z_t$) giving
the direction of the Sun in the observatory frame will thus be
related to the solar spherical coordinates($az_0$,$alt_0$) in the
altazimuthal system:
\begin{equation}
\begin{bmatrix}
x_{t} \\
y_{t} \\
z_{t}
\end{bmatrix}= M_{offsets}  \times
\begin{bmatrix}
\cos{alt_{0}}\cos{az_{0}} \\
\cos{alt_{0}}\sin{az_{0}} \\
\sin{alt_{0}}
\end{bmatrix}
\end{equation}

Substituting $M_{offset}$ with Equation (2) we get the following
expressions for those coordinates:
\begin{eqnarray}
\begin{cases}
x_{t} = \cos{(\alpha+az_{0})}\cos{\beta}\cos{alt_{0}}+\sin{\beta}\sin{alt_{0}} \\
y_{t} = (\cos{(\alpha+az_{0})}\sin{\beta}\sin{\gamma}+\sin{(\alpha+az_{0})}\cos{\gamma})\cos{alt_{0}} \\
\qquad- \cos{\beta}\sin{\gamma}\sin{alt_{0}} \\
z_{t} = (\sin{(\alpha+az_{0})}\sin{\gamma}-\cos{(\alpha+az_{0})}\sin{\beta}\cos{\gamma})\cos{alt_{0}} \\
\qquad+ \cos{\beta}\cos{\gamma}\sin{alt_{0}}
\end{cases}
\label{eq:euler}
\end{eqnarray}

These new Cartesian coordinates can then be converted to altitude
($alt_t$) and azimuth ($az_t$) angles relative to the tracker:
\begin{align}
\left\{
\begin{aligned}
\rho_{t}\,\,\, &= \sqrt{x_t^2+y_t^2} \\
alt_t &= \operatorname{atan_2}{(z_t,\rho_{t})} \\
az_t &= \operatorname{atan_2}{(y_t,x_{t})}\\
\end{aligned}\right.
\label{eq:eq5}
\end{align}

In the above equation, $atan_2(y,x)$, available in many programming
languages, stands for the argument of the complex number $x+iy$. It
is closely related to the arctangent of $y/x$  but it indicates
unambiguously the quadrant of this angle on the trigonometric
circle.

Determining Euler angles accurately by measurements is not easy. An
analytical method to estimate them is given in~\cite{Chong_2009}
which basically consists of recording the position of the tracker at
three different times and solving Equation~\eqref{eq:euler}. This is
appropriate for the studied case, \textit{i.e.},~a collector for
solar energy application installed outside with only one mirror and
no closed-loop control. With our considered two-mirror tracker,
which does not collect light but directs it toward a spectrometer,
other sources of misalignments appear. Indeed, the tracker is also
likely to be misaligned compared to the spectrometer, and mirror
themselves can be tilted. Other angles can be considered in
$M_{offsets}$ and is done in~\cite{Huster}. In practical
applications, despite the three Euler angles, the calculated mode is
likely able to reach the Sun within the FOV of the 4-quadrant diode.
With the closed-loop control it is easy to track the Sun
during a whole clear-sky day providing an operator
correctly sets the Sun tracker initially. Euler angles can then be
fitted using all the recorded positions of the mirrors during the
day. It has the advantage that other sources of misalignment are
included: even if only three angles are fitted which may not exactly
be the Euler angles, they minimize simultaneously the effects of all
offsets. We implement this method in Section~\ref{sec:application}.
This requires the closed-loop control of the tracker on the Sun
position.

\subsection{Ray Tracing in the Tracker}
\label{sec:active}

The photodiode signal indicates that the Sun beam is
tilted compared to the optical axis of the spectrometer. The
photodiode signals must be converted into angular movements of the
altitude and azimuth axes of the tracker to correct the
misalignment. If the photodiode was placed on the reference frame of
the mirror M2 this conversion would be straightforward, but due to
its position after the tracker it depends on the position of the
tracker mirrors. A trial-and-error method to correct
the misalignment is theoretically possible using analogue
electronics without a computer but a smoother tracking can be
achieved if the conversion is understood.

\begin{figure}[ht]
\centering
\includegraphics[width=.57\textwidth]{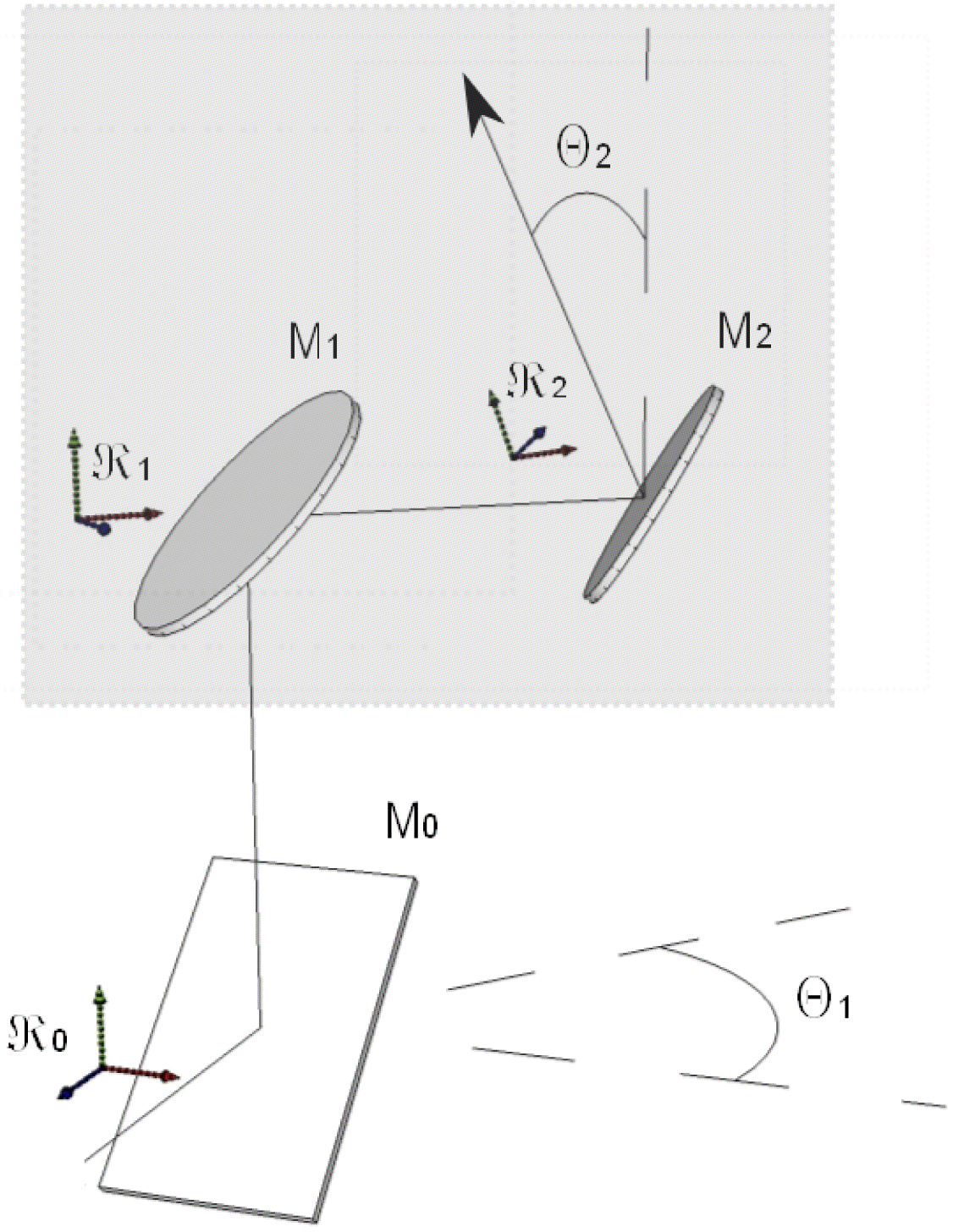}
\caption{The tracker mirrors and their rotation can be modeled as
rotation matrices in their reference frames, which apply to the beam
vector. Note that the $z$ and $y$ axes are the same
respectively for ($\Re_0$,$\Re_1$) and ($\Re_1$,$\Re_2$).}
\label{fig:trackos}
\end{figure}

The conversion can be expressed once again by a matrix, which
transforms in this case a vector hitting mirror M1 to a vector
pointing to a direction in the sky given by its altitude and
azimuth. It is the opposite of the light direction but is simpler to
figure out, and considering Fermat principle, yields the same
information.

The rotation of the two motorized stages can be
accounted for using rotation matrices as described in the previous
section.  The reflection on the two mirrors is modeled using another
matrix which takes the~form:
\begin{equation}
M = I - 2nn^T \label{eq:reflect}
\end{equation}

where $I$ is the identity matrix and $n$ the normal
vector to the mirror surface. At reference position, the mirror are
parallel and thus their normal is the same, given by the vector
($0$, $\frac{1}{\sqrt{2}}$,$-\frac{1}{\sqrt{2}}$). The
transformation matrix for the two mirror is thus the same, $M_R$,
which is, from Equation~\eqref{eq:reflect}:
\begin{equation}
M_R =
\begin{bmatrix}
1 & 0 & 0 \\
0 & 0 & 1 \\
0 & 1 & 0 \end{bmatrix}
\end{equation}

Figure~\ref{fig:trackos} presents the tracker pointing to an azimuth
$\theta_1$ and a zenith angle of $\theta_2$. The reference frames
$\Re_1$ and $\Re_2$ are respectively attached to the mirrors $M_1$
and $M_2$, with the $x$ axes in the direction of their small
semi-axes and the $y$ axes along the line joining the two mirrors.
The optical system inside the frame, with only mirrors $M_1$ and
$M_2$, can be expressed as a transformation whose matrix
$M_{tracker}$ is:
\begin{equation}
M_{tracker} =R_z(\theta_1) \times R_y(\theta_2) \times M_R \times
R_y(-\theta_2) \times M_R \times  R_z(-\theta_1)
\end{equation}
The above formula is derived
as follow: (a) the reflection on $M_2$ ($M_R$) is expressed
in the reference frame of $\Re_1$ with a change of basis
involving $R_y(-\theta_2)$; (b) this product of three matrices
is multiplied on its right side by the preceding (seen from
the spectrometer) reflection on $M_1$($M_R$); (c) another change
of basis is performed to express the transformation in $\Re_0$,
involving $R_z(-\theta_1)$. %
\textit{i.e.},

$M_{tracker} = $

\begin{align}
\begin{bmatrix}
\cos{\theta_1} & -\sin{\theta_1} & 0 \\
\sin{\theta_1} & \cos{\theta_1} & 0 \\
0 & 0 & 1 \end{bmatrix}  \times
\begin{bmatrix}
\cos{\theta_2} & 0 & \sin{\theta_2} \\
0 & 1 & 0 \\
-\sin{\theta_2} & 0 & \cos{\theta_2} \end{bmatrix} \times
\begin{bmatrix}
1 & 0 & 0 \\
0 & 0 & 1 \\
0 & 1 & 0 \end{bmatrix} \times \nonumber \\
\begin{bmatrix}
\cos{\theta_2} & 0 & -\sin{\theta_2} \\
0 & 1 & 0 \\
\sin{\theta_2} & 0 & \cos{\theta_2} \end{bmatrix}  \times
\begin{bmatrix}
1 & 0 & 0 \\
0 & 0 & 1 \\
0 & 1 & 0 \end{bmatrix}  \times
\begin{bmatrix}
\cos{\theta_1} & \sin{\theta_1} & 0 \\
-\sin{\theta_1} & \cos{\theta_1} & 0 \\
0 & 0 & 1  \end{bmatrix} 
\end{align}

\vspace{5pt}

Developing the matrix product yields the matrix of the tracker
optical system as a function of the tracker position
($\theta_1$,$\theta_2$):

\begin{align}
\lefteqn{M_{tracker} =  }  \\ &
\left[ \begin{smallmatrix}
\cos{\theta1}\cos{\theta2}\cos({\theta_1-\theta_2})+\sin{\theta_1}\sin({\theta_1-\theta_2})
&
\cos{\theta_1}\cos{\theta_2}\sin({\theta_1-\theta_2})-\sin{\theta_1}\cos({\theta_1-\theta_2})
&
\cos{\theta_1}\sin{\theta_2} \nonumber \\
\sin{\theta_1}\cos{\theta_2}\cos({\theta_1-\theta_2})-\cos{\theta_1}\sin({\theta_1-\theta_2})
&
\cos{\theta_1}\cos({\theta_1-\theta_2})+\sin{\theta_1}\cos{\theta_2}\sin({\theta_1-\theta_2})
&
\sin{\theta_1}\sin{\theta_2} \nonumber \\
-\sin{\theta_2}\cos{\theta_1-\theta_2} &
-\sin{\theta_2}\sin{\theta_1-\theta_2} & \cos{\theta_2}
\end{smallmatrix} \right]  \nonumber \\  \nonumber
\end{align}

\vspace{5pt}

\begin{figure}[ht]
\centering
\includegraphics[width=.26\textwidth]{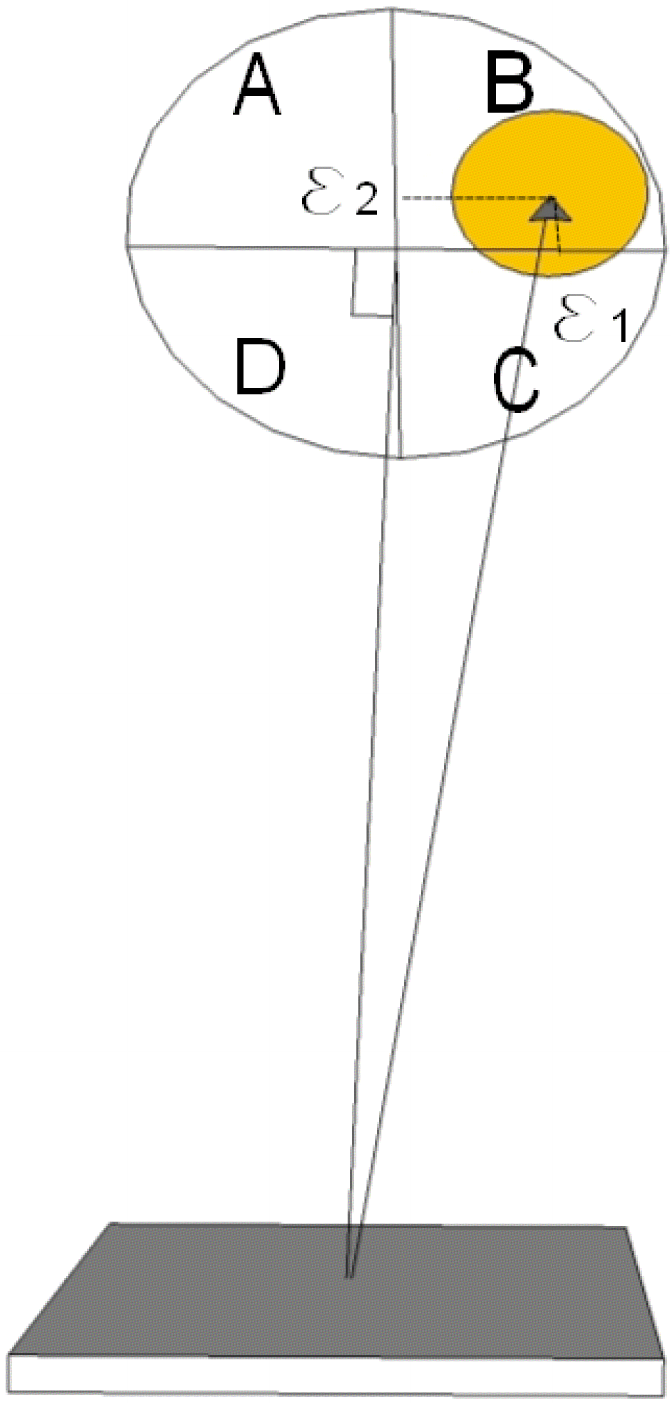}
\caption{Sun spot hitting the quadrant, not to scale.}
\label{fig:sunspot}
\end{figure}

The transformation expressed by $M_{tracker}$ can now be applied to
a vector corresponding to the Sun light beam direction on the
spectrometer side of the tracker. It will lead to the position of
the Sun in Cartesian coordinates. The vector is built from the
4 diode signals (VA,VB,VC,VD), as represented in
Figure~\ref{fig:sunspot}. Basically an offset position
($\varepsilon_1$,$\varepsilon_2$) is computed for the Sun spot on
the diode plane compared to its center by:
\begin{eqnarray}
\begin{cases}
\varepsilon_1 = (V_B + V_C) - (V_A + V_D) \\
\varepsilon_2 = (V_A + V_B) - (V_C + V_D)
\end{cases}
\end{eqnarray}

\vspace{5pt}

The spot offset ($\varepsilon_1$,$\varepsilon_2$) defines 2
coordinates of the beam vector. The last one, $\Lambda$, should
represent the distance from the diode to mirror M1. Multiplying
$M_{tracker}$ by the quadrant vector
($\varepsilon_1$,$\varepsilon_2$,$\Lambda$) would yield accurate Sun
angles after conversion to spherical coordinates, but is not
practically possible with a diode, contrary to an imaging sensor.
The calculated position ($x_s$,$y_s$,$z_s$) hereafter is thus not
absolute but is sufficient to get the sign of the rotations to apply
on the axes. $\Lambda$ can be chosen arbitrarily as
long as its absolute value is large enough compared to
$\varepsilon_1$ and $\varepsilon_2$. The solar pseudo-coordinates
are then:
\begin{equation}
\begin{bmatrix}
x_{s} \\
y_{s} \\
z_{s}
\end{bmatrix}= M_{tracker}  \times
\begin{bmatrix}
\varepsilon_1 \\
\varepsilon_2 \\
\Lambda
\end{bmatrix}
\label{eq:quadvector}
\end{equation}

In practice, the quadrant vector may differ from
($\varepsilon_1$,$\varepsilon_2$,$\Lambda$) due to
reflections such as on the mirrors M0 and M4q on Figure~\ref{fig:principle}, 
necessary to deviate a part of the beam to the 4-quadrant
photodiode. Defining the position vector thus requires to pay
attention to the optical path from M1 to the photodiode. In section~\ref{sec:application}, we explain
how we deal with the problem in our particular case.

Developing Equation~\eqref{eq:quadvector} yields:
\begin{eqnarray}
\begin{cases}
x_s = (\cos{\theta_1}\cos{\theta_2}\sin{(\theta_1-\theta_2)}-\sin{\theta_1}\cos{(\theta_1-\theta_2)})\varepsilon_2 \\
\qquad{}+(\sin{\theta_1}\sin{(\theta_1-\theta_2)}+\cos{\theta_1}\cos{\theta_2}\cos{(\theta_1-\theta_2)})\varepsilon_1 +\Lambda\cos{\theta_1}\sin{\theta_2} \\
y_s = (\sin{\theta_1}\cos{b}\sin{(\theta_1-\theta_2)}+\cos{\theta_1}\cos{(\theta_1-\theta_2)})\varepsilon_2\\
\qquad{}+(\sin{\theta_1}\cos{\theta_2}\cos{(\theta_1-\theta_2)}-\cos{\theta_1}\sin{(\theta_1-\theta_2)})\varepsilon_1 +\Lambda\sin{\theta_1}\sin{\theta_2} \\
z_s =
-\sin{\theta_2}\sin{(\theta_1-\theta_2)}\varepsilon_2-\sin{\theta_2}\cos{(\theta_1-\theta_2)}\varepsilon_1+\Lambda\cos{\theta_2}
\end{cases}
\end{eqnarray}

It is then possible to calculate roughly an
altitude($\theta_{2S}$) and azimuth($\theta_{1S}$) for the Sun applying the
Cartesian to spherical coordinates conversion (Equation~\eqref{eq:eq5}). 
This position is approximate and relative to the tracker since it
does not take into account the Euler angles described in the last
section, but what matters are the signs of the
differences between these calculated values and the current altitude
and azimuth relative to the tracker, defined by $\theta_1$ and
$\theta_2$. The angular corrections to apply on the two axes are
then:
\begin{eqnarray}
\begin{cases}
d_{\theta1} = sgn(\theta_{1S}-\theta_1)k_1 \\
d_{\theta2} = sgn(\theta_{2S}-\theta_2)k_2
\end{cases}
\label{eq:dtheta}
\end{eqnarray}
where $k_1$ and $k_2$ are the tracking angle steps that should be
small to have a smooth tracking, yet large enough for the mechanical
resolution of the rotation stages and the apparent movement of the
Sun. The azimuth changes for instance at a rate of 15$^{\circ}$ per
hour, assuming 1 second between the steps, $k_1$ should not be under
0.004$^{\circ}$.

\section{Automation Issues}\label{sec:sec3}

From a control theory perspective, the altazimuthal tracker and its
feedback is a non-linear \mbox{multi-input} multi-output (MIMO)
system. Indeed, two outputs defining the pointing direction
($\theta_1$ and $\theta_2$ ) are controlled by two inputs,
\textit{i.e.},~the coordinates of the Sun spot on the
photodiode($\varepsilon_1$ and $\varepsilon_2$), and the
relationship between the inputs and the outputs varies with the
position of the tracker. However, having modeled this relationship
in the previous section, it is possible to change the feedback
scheme while tracking. In control theory, this is an example of
adaptive control.

\begin{figure}[ht]
\centering
\includegraphics[width=.95\textwidth]{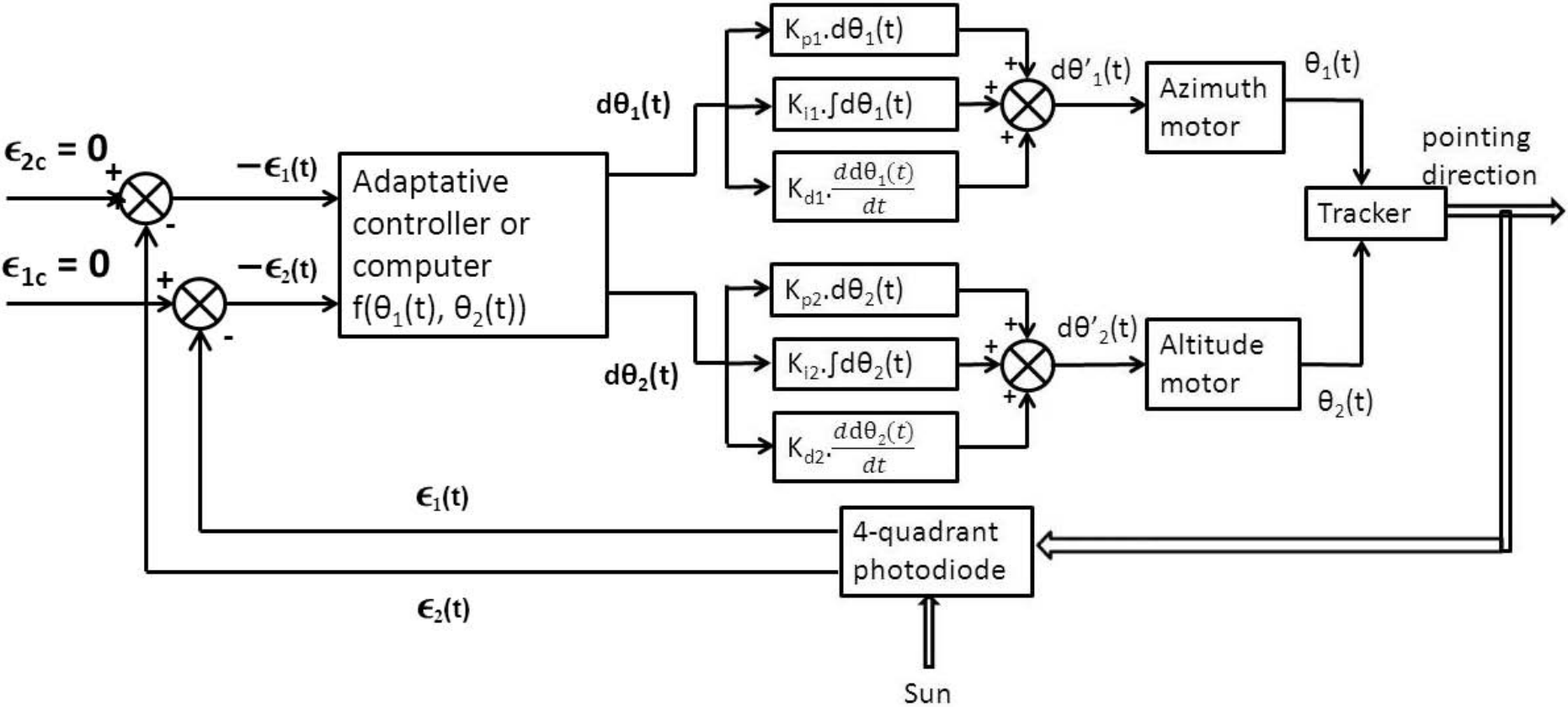}
\caption{Control loop for an
altazimuthal tracker.} \label{fig:control}
\end{figure}

The correction of the azimuth and altitude angles discussed earlier
only takes into account the current error, \textit{i.e.},~the tilt
of the solar beam compared to the optical axis of the spectrometer.
This is very coarse and can lead to oscillations. A proper feedback
loop includes the derivative and integral of the error relative to
the time as well, respectively to reduce the overshoot and the
residual part of the errors. This involves to tune the three
parameters of a proportional\textendash integral\textendash
derivative (PID) controller. Considering the two outputs, the setup
needs two PID controllers.

Figure ~\ref{fig:control} suggests a feedback loop for the tracker.
Note that $\varepsilon_{1c}$ and $\varepsilon_{2c}$ are null if the
photodiode is correctly aligned compared to the spectrometer optical
axis. The adaptive control algorithm on the figure
($f(\theta_1(t),\theta_2(t))$) originates from the formula derived
in the previous section. There are 6 parameters to tune to optimize
the feedback, which corresponds to the proportional, derivative and
integral terms of the two PID controllers. Several methods exist to
optimize PID parameters, with different complexity. We think the
simple Ziegler--Nichols method could be appropriate. It consists in
setting Ki and Kd to 0 and increasing Kp from 0 to the value Kpc
with which oscillations occur at constant amplitude, with a period
Tc. Three good values for Kp, Ki and Kd can then be derived as:
\begin{align}
\left\{
\begin{aligned}
K_p &= 0.6K_c \\
K_i &= \frac{2K_p}{T_c} \\
K_d &= \frac{K_p T_c}{8} \\
\end{aligned}
\right.
\end{align}

\vspace{-6pt}

Considering the latitude of the Reunion Island observatory where the
tracker is installed (20.9$^{\circ}$S) it is worth considering an
issue occurring with the altazimuthal geometry, \textit{i.e.},~the
singularity at zenith. As the altitude gets closer to zenith, the
azimuth rotation gets more and more difficult to control. At Reunion
Island, the maximum altitude is reached around November the 26th and
January the 16th. Around this date, the measurement dead time can
reach one hour. To our knowledge, there is no good solution to solve
the problem. Nevertheless, to limit the measurement dead time around
noon, we propose to implement in the adaptive controller another
mode starting when the altitude is too high and the problem happens:
shifting the azimuth mode from feedback controlled to calculated
mode. The elevation is still controlled by the photodiode. This may
not be accurate enough to keep the Sun spot in the spectrometer's
iris, but it would at least avoid possible incoherent movement of
the tracker.

\section{Application for a FTIR Measurement Station}
\label{sec:application}

Our group has been doing FTIR measurements at Reunion Island for
several years (\cite{demaziere_2008, senten_2008, Vigouroux-2009}).
The place is interesting since atmospheric measurements are sparse
in the tropical and subtropical regions. Aiming at long-term
monitoring and cost-effectiveness, a station at Saint-Denis
(20.9$^{\circ}$S, 55.5$^{\circ}$E, 50~m a.s.l.) has been
automated~\cite{neefs_2007}, which includes Sun tracking,
meteorological logging and FTIR measurements with a Bruker 120M. The
solar tracker currently used was developed at Denver University.
Since September 2009, this station is officially part of the NDACC
network and in Spring 2012 it will move to the new Maido Observatory
(21.1$^{\circ}$S, 55.4$^{\circ}$E, 2200 m a.s.l.).

\begin{figure}[ht]
\centering
\includegraphics[width=.63\textwidth]{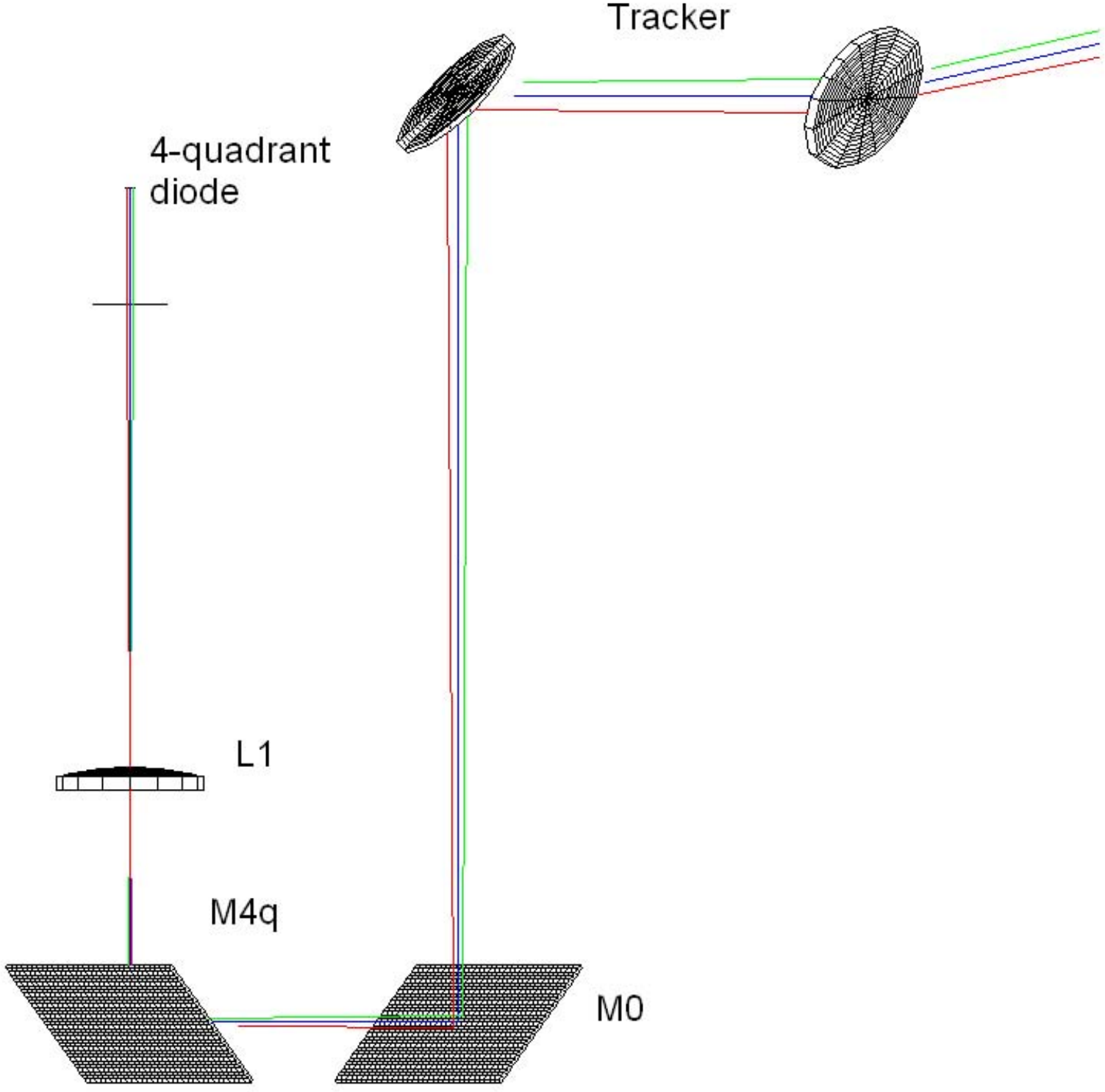}
\caption{Optical paths of our particular set-up from the tracker to
the 4-quadrant diode (not to scale).} \label{fig:opath}
\end{figure}

A second FTIR station has been installed in Saint-Denis in September
2012. This station, which is also automated, is
based on a Bruker 125 HR spectrometer, more appropriate to measure
CO$_2$ atmospheric loading, in the framework of the new Total Carbon
Column Observing Network (TCCON). The geometry of the Sun tracker is
altazimuthal. It was built at our institute and used to validate the
methods described in the last section.

This new solar tracker uses a Newport RV-160 rotation stage for the
azimuth rotation and a Vexta stepping motor with a
gear box for the altitude. Both rotations are driven by a Newport
XPS controller, linked to the controlling PC. The tracker mirrors
are elliptical with a 10 cm minor axis. The photodiode setup was
purchased from Bruker with the spectrometer and is installed at the
input window of the spectrometer. It consists of a 1 cm mirror which
reflects a small portion of the incoming light to a 18 cm focal
length lens which focuses the beam onto the 4-quadrant photodiode.
The optical path is shown in Figure~\ref{fig:opath}. The
FOV of the 4-quadrant photodiode is 20 mrad (see
Section~\ref{sec:fov}). The algorithm used to compute the ephemeris
is the one given in appendix. During operation, the mirrors
positions are refreshed every half second according to the
calculated position or to the signal on the 4-quadrant photodiode
using the methods described in Section~\ref{sec:active}. From
Figure~\ref{fig:opath}, it is clear than the beam from the tracker
undergoes two orthogonal reflections before
hitting the photodiode, which we take into account multiplying $Mtracker$ on its rigth side by the corresponding factors derived from Equation~\eqref{eq:reflect}.

\begin{figure}[ht]
\centering
\includegraphics[width=\textwidth]{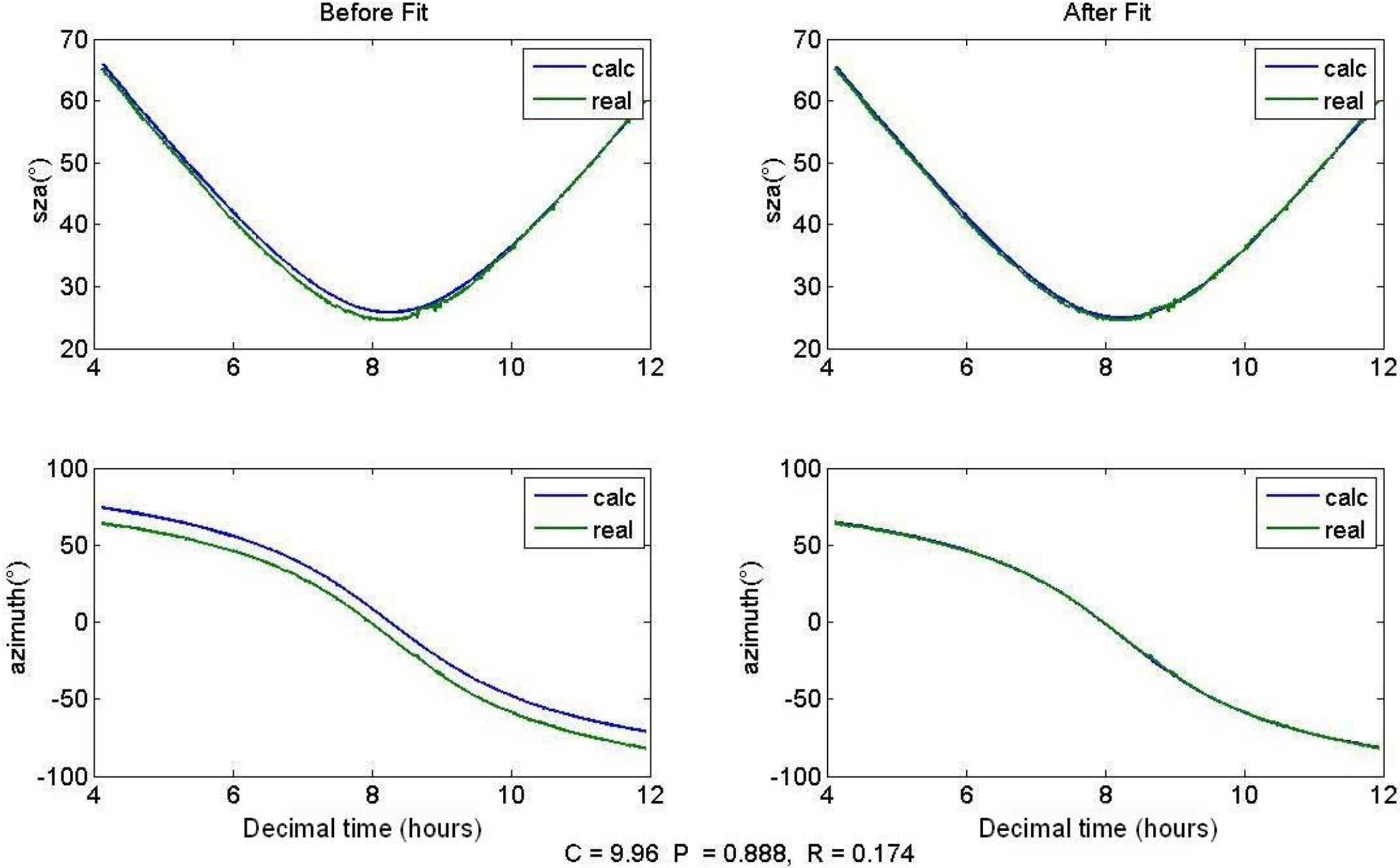}
\caption{Fit of the Euler Angles to take into account the alignment
offsets in the calculation mode. The track was performed on 12
September 2011.} \label{fig:fit}
\end{figure}

Figure~\ref{fig:fit} shows a fit of the Euler Angles. The track was
performed on 12 September 2011 in \mbox{Saint-Denis}. The left panel
shows the calculated mirror positions neglecting Euler Angles,
together with the actual ones when the active tracker was
operational. Weather was clear-sky and enabled to record a long Sun
path, demonstrating the capacity of the active tracking algorithm
described in Section~\ref{sec:active}. Around one hundred points
were extracted from the log file of the tracker position and used to
fit the Euler Angles with an unconstrained nonlinear minimization.
The three angles $\alpha$,$\beta$ and $\gamma$ described in
Section~\ref{sec:euler} were respectively estimated to be
9.96$^{\circ}$, 0.888$^{\circ}$ and 0.174$^{\circ}$. The right panel
shows how these fitted angles improve the calculated Sun position.
The maximum offset between calculated and actual position is now
0.5$^{\circ}$, \textit{i.e.},~9 mrad, which is under the 20 mrad of
the photodiode's FOV. Providing this accuracy in the calculated
mode, the tracking system is able to set the Sun's image onto the
4-quadrant photodiode and then start the active tracking without the
need of an operator.

\section{Conclusions}

We have derived the geometrical formulas needed to track the Sun
with a kind of altazimuthal tracker widely used in atmospheric
remote sensing. The setup is based on two rotating 45$^{\circ}$
mirrors facing each other and a 4-quadrant photodiode involved in a
closed-loop control of the tracker. After discussing the required
accuracy for the calculated mode and calculating the FOV of the
sensor, we described how to take into account and estimate the Euler
angles, representing the orientation of the tracker compared to the
ground. These sections can actually be applied to other tracking
setups. On the other hand, even if the method is general, the
formula for the active tracking depends strongly on the optical
configuration and may not be used for other trackers' geometries. We
have proposed a control loop with PID to achieve a smooth tracking
while reducing overshoot and the residual part of the error.
Finally, we have tested the formulas with a custom-built solar
tracker that has been installed together with a FTIR spectrometer at
Reunion Island in September 2011.

Among the future work will be the improvement of the tracking
smoothness, and particularly the tuning of the six parameters of the
PID controllers. We will also implement the solution presented in
Section~\ref{sec:sec3} and check whether the measurement dead time
can be reduced.

A characteristic of the Maido observatory is the very regular cloud
cycle. At noon, the clouds reach the observatory almost every day.
This is very convenient for clouds studies but less for solar
occultation trace gases measurements. On the other hand, the nights
are so clear up there that the observatory was first supposed to be
dedicated to astronomical research. This is thus a good place to try
Moon tracking and we plan to work on that in the future.

\section*{Acknowledgements}

This work was funded by the Belgian Science Policy (BELSPO). The
authors wish to thank Thomas Blumenstock for his advices and for
having sent him the work of M. Huster. They also thank Filip Desmet,
Bart Dils and S\'ebastien Henrotin for useful discussions.

\bibliographystyle{ieeetr}
\bibliography{tracksensors}

\begin{thebibliography}{10}

\bibitem{barret2002}
B.~Barret, M.~{De Mazi\`{e}re}, and P.~Demoulin, ``Retrieval and
  characterization of ozone profiles from solar infrared spectra at the
  jungfraujoch,'' {\em J. Geophys. Res.}, vol.~107, no.~D24, p.~15 PP., 2002.

\bibitem{demaziere_2005}
M.~{De Mazi\`{e}re}, C.~Vigouroux, T.~Gardiner, M.~Coleman, P.~Woods,
  K.~Ellingsen, M.~Gauss, I.~Isaksen, T.~Blumenstock, F.~Hase, I.~Kramer,
  C.~Camy-peyret, P.~Chelin, E.~Mahieu, P.~Demoulin, P.~Duchatelet,
  J.~Mellqvist, A.~Strandberg, V.~Velazco, J.~Notholt, R.~Sussmann, W.~Stremme,
  and A.~Rockmann, ``The exploitation of ground-based fourier transform
  infrared observations for the evaluation of tropospheric trends of greenhouse
  gases over europe,'' {\em Environm. Sci.}, vol.~2, pp.~283--293, 2005.

\bibitem{mvr_1998}
M.~V. Roozendael, P.~Peeters, H.~K. Roscoe, H.~D. Backer, A.~E. Jones,
  L.~Bartlett, G.~Vaughan, F.~Goutail, J.~P. Pommereau, E.~Kyro, C.~Wahlstrom,
  G.~Braathen, and P.~C. Simon, ``Validation of {Ground-Based} visible
  measurements of total ozone by comparison with dobson and brewer
  spectrophotometers,'' {\em J. Atmos. Chem.}, vol.~29, pp.~55--83, 1998.

\bibitem{cageao_2001}
R.~P. Cageao, J.~Blavier, J.~P. {McGuire}, Y.~Jiang, V.~Nemtchinov, F.~P.
  Mills, and S.~P. Sander, ``{High-Resolution} {Fourier-Transform}
  {Ultraviolet-Visible} spectrometer for the measurement of atmospheric trace
  species: Application to {OH},'' {\em Appl. Optics}, vol.~40, pp.~2024--2030,
  2001.

\bibitem{wang-2010}
S.~Wang, T.~J. Pongetti, S.~P. Sander, E.~Spinei, G.~H. Mount, A.~Cede, and
  J.~Herman, ``Direct sun measurements of {NO2} column abundances from table
  mountain, california: Intercomparison of low- and high-resolution
  spectrometers,'' {\em J. Geophys. Res.}, vol.~115, no.~D13305, p.~16 PP.,
  2010.

\bibitem{Spinei2011}
E.~Spinei and G.~H. Mount, ``O2-o2 absorption cross section derived from direct
  sun measurements at different locations,'' in {\em OMI Science Team Meeting
  nr. 15}, 2010.

\bibitem{galle_1999}
B.~Galle, J.~Mellqvist, D.~W. Arlander, I.~Floisand, M.~P. Chipperfield, and
  A.~M. Lee, ``Ground based {FTIR} measurements of stratospheric species from
  harestua, norway during sesame and comparison with models,'' 1999.

\bibitem{wiacek_2007}
A.~Wiacek, J.~R. Taylor, K.~Strong, R.~Saari, T.~E. Kerzenmacher, N.~B. Jones,
  and D.~W.~T. Griffith, ``Ground-based solar absorption {FTIR} spectroscopy:
  Characterization of retrievals and first results from a novel optical design
  instrument at a new {NDACC} complementary station,'' {\em J. Atmos. Ocean.
  Technol.}, vol.~24, pp.~432--448, Mar. 2007.

\bibitem{neefs_2007}
E.~Neefs, M.~{De Maziere}, F.~Scolas, C.~Hermans, and T.~Hawat, ``{BARCOS,} an
  automation and remote control system for atmospheric observations with a
  bruker interferometer,'' {\em Rev. Sci. Instrum.}, vol.~78, no.~3, p.~8 PP.,
  2007.

\bibitem{Huster}
M.~Huster, ``Bau eines automatischen sonnenverfolgers für bodengebundene
  ir-absorptionsmessungen,'' Master's thesis, IMK, 1998.

\bibitem{washenfelder_2006}
R.~A. Washenfelder, G.~C. Toon, J.~Blavier, Z.~Yang, N.~T. Allen, P.~O.
  Wennberg, S.~A. Vay, D.~M. Matross, and B.~C. Daube, ``Carbon dioxide column
  abundances at the wisconsin tall tower site,'' {\em J. Geophys. Res.},
  vol.~111, no.~D22305, p.~16 PP., 2006.

\bibitem{merlaud2006}
A.~Merlaud, ``Development of a solar tracker for monitoring of atmospheric
  gases at harestua observatory,'' tech. rep., Chalmers Insitute of Technology,
  2006.

\bibitem{merlaud2004}
A.~Merlaud, ``Development of solar tracker for studies of volcanic gas
  emissions,'' Master's thesis, ENSPG, 2004.

\bibitem{Cordenier}
A.~Cordenier, ``Syst\`{e}me num\'{e}rique de r\'{e}gulation en position de
  miroirs destin\'{e}s au suivi du rayonnement solaire,'' Master's thesis,
  ECAM, 2004.

\bibitem{geibel-2010}
M.~C. Geibel, C.~Gerbig, and D.~G. Feist, ``A new fully automated {FTIR} system
  for total column measurements of greenhouse gases,'' {\em Atmos. Meas.
  Tech.}, vol.~3, pp.~1363--1375, 2010.

\bibitem{Gisi_2011}
M.~Gisi, F.~Hase, S.~Dohe, and T.~Blumenstock, ``Camtracker: a new camera
  controlled high precision solar tracker system for ftir-spectrometers,'' {\em
  Atmospheric Measurement Techniques}, vol.~4, no.~1, pp.~47--54, 2011.

\bibitem{Chong_2009}
K.~Chong and C.~Wong, ``General formula for on-axis sun-tracking system and its
  application in improving tracking accuracy of solar collector,'' {\em Solar
  Energy}, vol.~83, no.~3, pp.~298--305, 2009.

\bibitem{Chong_2009b}
K.~K. Chong, C.~W. Wong, F.~L. Siaw, T.~K. Yew, S.~S. Ng, M.~S. Liang, Y.~S.
  Lim, and S.~L. Lau, ``Integration of an on-axis general sun-tracking formula
  in the algorithm of an open-loop sun-tracking system,'' {\em Sensors},
  vol.~9, pp.~7849--7865, 2009.

\bibitem{guo_2011}
M.~Guo, Z.~Wang, J.~Zhang, F.~Sun, and X.~Zhang, ``Accurate altitude–azimuth
  tracking angle formulas for a heliostat with mirror–pivot offset and other
  fixed geometrical errors,'' {\em Sol. Energy}, vol.~85, pp.~1091--1100, May
  2011.

\bibitem{wei_2011}
X.~Wei, Z.~Lu, W.~Yu, H.~Zhang, and Z.~Wang, ``Tracking and ray tracing
  equations for the target-aligned heliostat for solar tower power plants,''
  {\em Renewable Energy}, vol.~36, pp.~2687--2693, Oct. 2011.

\bibitem{meeus_astronomical_1998}
J.~Meeus, {\em Astronomical Algorithms}.
\newblock Atlantic Books, 2nd illustrated edition~ed., 1998.

\bibitem{reda_solar_2008}
I.~Reda and A.~Andreas, ``Solar position algorithm for solar radiation
  applications,'' {\em Sol. Energy}, vol.~76, no.~5, pp.~577--589, 2004.

\bibitem{Ciddor96}
P.~E. Ciddor, ``Refractive index of air: new equations for the visible and near
  infrared,'' {\em Appl. Optics}, vol.~35, pp.~1566--1573, 1996.

\bibitem{demaziere_2008}
M.~{De Mazi\`{e}re}, C.~Vigouroux, P.~F. Bernath, P.~Baron, T.~Blumenstock,
  C.~Boone, C.~Brogniez, V.~Catoire, M.~Coffey, P.~Duchatelet, D.~Griffith,
  J.~Hannigan, Y.~Kasai, I.~Kramer, N.~Jones, E.~Mahieu, G.~L. Manney,
  C.~Piccolo, C.~Randall, C.~Robert, C.~Senten, K.~Strong, J.~Taylor,
  C.~T\'{e}tard, K.~A. Walker, and S.~Wood, ``Validation of {ACE-FTS} v2.2
  methane profiles from the upper troposphere to the lower mesosphere,'' {\em
  Atmos. Chem. Phys.}, vol.~8, no.~9, pp.~2421--2435, 2008.

\bibitem{senten_2008}
C.~Senten, M.~{De Mazi\`{e}re}, B.~Dils, C.~Hermans, M.~Kruglanski, E.~Neefs,
  F.~Scolas, A.~C. Vandaele, G.~Vanhaelewyn, C.~Vigouroux, M.~Carleer, P.~F.
  Coheur, S.~Fally, B.~Barret, J.~L. Baray, R.~Delmas, J.~Leveau, J.~M.
  Metzger, E.~Mahieu, C.~Boone, K.~A. Walker, P.~F. Bernath, and K.~Strong,
  ``Technical note: New ground-based {FTIR} measurements at ile de la
  r\'{e}union: observations, error analysis, and comparisons with independent
  data,'' {\em Atmos. Chem. Phys.}, vol.~8, pp.~3483--3508, 2008.

\bibitem{Vigouroux-2009}
C.~Vigouroux, F.~Hendrick, T.~Stavrakou, B.~Dils, I.~D. Smedt, C.~Hermans,
  A.~Merlaud, F.~Scolas, C.~Senten, G.~Vanhaelewyn, S.~Fally, M.~Carleer,
  J.~Metzger, J.~M\"{u}ller, M.~V. Roozendael, and M.~D. Mazi\`{e}re,
  ``Ground-based {FTIR} and {MAX-DOAS} observations of formaldehyde at
  r\'{e}union island and comparisons with satellite and model data,'' {\em
  Atmos. Chem. Phys.}, vol.~9, pp.~9523--9544, 2009.

\bibitem{Iqbal_1983}
M.~Iqbal, {\em An introduction to solar radiation}.
\newblock Academic Press, 1983.

\end{thebibliography}


\appendix
\section{Equations for Solar Ephemeris}
\makeatletter \@addtoreset{figure}{section} \makeatother
\setcounter{figure}{0}
\renewcommand{\thefigure}{\bfseries A\arabic{figure}}

For the sake of completeness, we reproduce here the simple algorithm
we use to calculate the solar coordinates given a date and a
position, taken from~\cite{Iqbal_1983}. It should be accurate enough for most tracking
purposes.

We first compute the fractional year ($\gamma$) in radians:
\begin{equation}
\gamma  = \frac{2\pi}{365}*(JD-1+\frac{T-12}{24}) \label{eq:xdef1}
\end{equation}
where $JD$ stands for the Julian day and $T$ for the UTC decimal
time.

Then we derive the equation of time ($\Delta t$) in minutes:
\begin{align}
\begin{split}
\Delta t  =  229.18*(0.000075&+ 0.001868\cos{\gamma} - 0.032077\sin{\gamma}  \\
  {}  -0.014615\cos{2\gamma} &- 0.040849\sin{2\gamma})
\end{split}
\end{align}

The equation of time represents the difference between apparent
solar time and mean solar time, which can be as large as 16 min. It
is due to the obliquity of the ecliptic and the elliptical form of
the earth orbit.

From the fractional year, we also get the solar declination
($\delta_{\odot}$) in radians:
\begin{align}
\begin{split}
\delta_{\odot}  = \,&  0.006918 - 0.399912\cos{\gamma} + 0.070257\sin{\gamma} -  0.006758\cos{2\gamma}  \\
  & {} +  0.000907\sin{2\gamma} - 0.002697\cos{3\gamma} + 0.00148\sin{3\gamma}
\end{split} \label{eq:xdef2}
\end{align}

The declination is the equivalent of the latitude on the celestial
sphere.

The offset $T_{off}$ (in minutes) between the UTC time and the true
solar time depends on the longitude (in degrees East) and is:
\begin{equation}
T_{off} = \Delta t-4*longitude \label{eq:xdef3}
\end{equation}

The true solar time ($tst$, in minutes) is then:
\begin{equation}
tst = hour*60+min+sec/60+T_{off} \label{eq:xdef4}
\end{equation}
where $hour$, $min$ and $sec$ are the components of the UTC time.

The solar hour angle, in degrees, comes from the true solar time as:
\begin{equation}
ha = \frac{tst}{4}-180 \label{eq:xdef5}
\end{equation}

For a given latitude, the hour angle and the declination are
converted to horizontal coordinates, \textit{i.e.},~solar zenith
angle ($\phi$) and azimuth ($\theta$, from south positive eastwards)
as:
\begin{align}
\phi &= \arccos{(\sin{lat}\sin{\delta_{\odot}}+\cos{lat}\cos{\delta_{\odot}}\cos{ha})}\\
%
\theta &= -
\operatorname{atan_2}({\sin{ha}\cos{\delta_{\odot}},\cos{ha}\sin{lat}\cos{\delta_{\odot}}-\cos{lat}\sin{\delta_{\odot}}})
\end{align}

Finally, the effect of refraction in arcminutes can
be approximated using S\ae mundsson's
formula~\cite{meeus_astronomical_1998}:
\begin{equation}
R = \frac{1.02}{\tan{h + \frac{10.3}{h+5.1}}}
\end{equation}

where $h$ is the unrefracted altitude in degree.

\begin{figure}[ht]
\centering
\includegraphics[width=130mm]{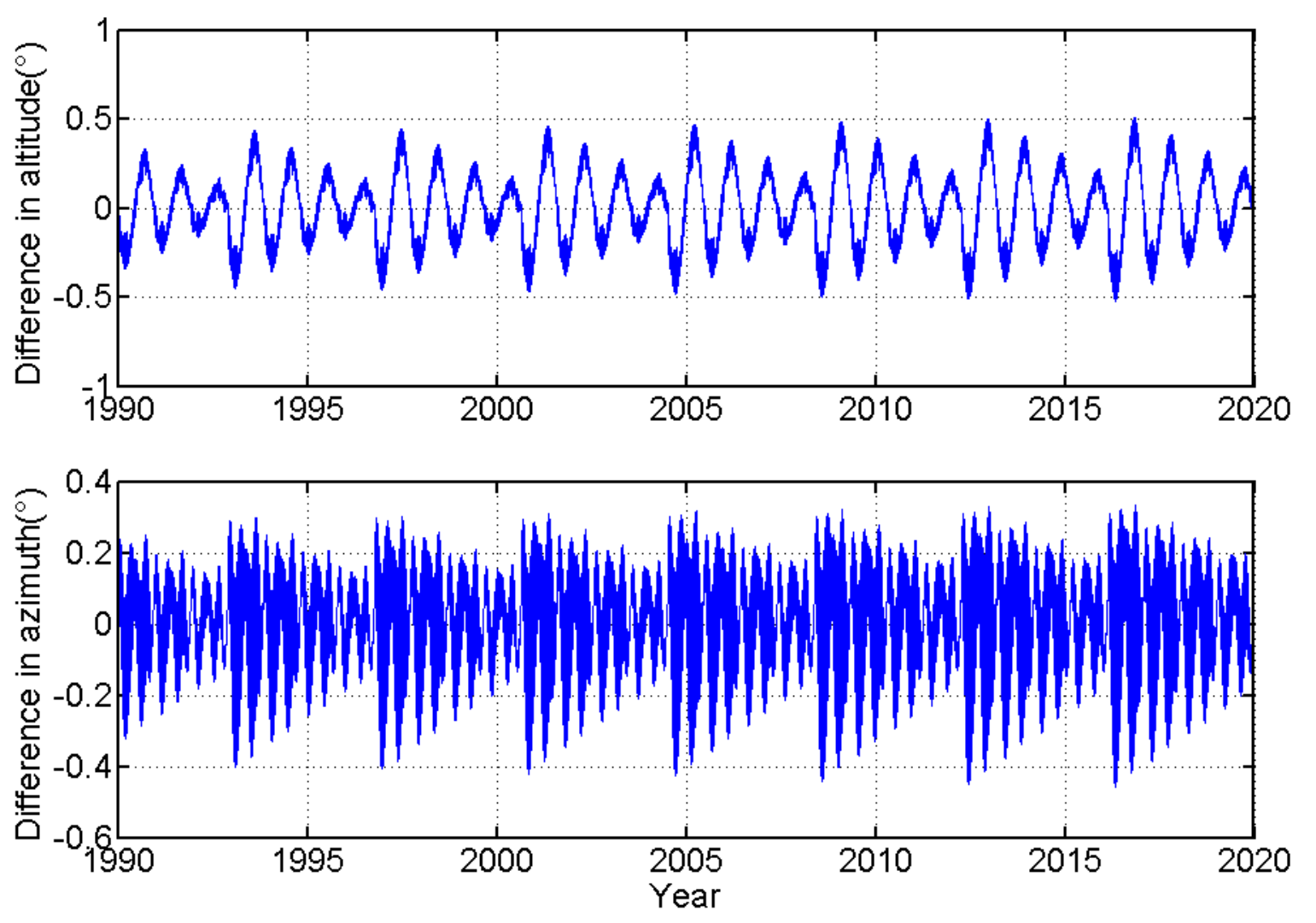}
\caption{Comparison between the formulas reproduced
in appendix and JPL ephemerides.} \label{fig:opathA}
\end{figure}

Figure \ref{fig:opathA} shows a comparison between the algorithm presented below and the Jet Propulsion Laboratory HORIZONS \footnote{http://ssd.jpl.nasa.gov/?horizons} ephemeris calculator. We compare the altitude and azimuth of the Sun seen from Brussels (50.85$^{\circ}$N,4.35$^{\circ}$E) over a 30-year period, between 1990 and 2020. Differences in altitude and azimuth are within 0.5$^{\circ}$.

\end{document}